# Distributed Multicell Beamforming Design Approaching Pareto Boundary with Max-Min Fairness

Yongming Huang, *Member, IEEE,* Gan Zheng, *Member, IEEE,* Mats Bengtsson, *Senior Member, IEEE*, Kai-Kit Wong, *Senior Member, IEEE,* Luxi Yang, *Member, IEEE,* and Björn Ottersten, *Fellow, IEEE*

*Abstract*—This paper addresses coordinated downlink beamforming optimization in multicell time-division duplex (TDD) systems where a small number of parameters are exchanged between cells but with no data sharing. With the goal to reach the point on the Pareto boundary with max-min rate fairness, we first develop a two-step centralized optimization algorithm to design the joint beamforming vectors. This algorithm can achieve a further sum-rate improvement over the max-min optimal performance, and is shown to guarantee max-min Pareto optimality for scenarios with two base stations (BSs) each serving a single user. To realize a distributed solution with limited intercell communication, we then propose an iterative algorithm by exploiting an approximate uplink-downlink duality, in which only a small number of positive scalars are shared between cells in each iteration. Simulation results show that the proposed distributed solution achieves a fairness rate performance close to the centralized algorithm while it has a better sum-rate performance, and demonstrates a better tradeoff between sum-rate and fairness than the Nash Bargaining solution especially at high signal-to-noise ratio.

## I. INTRODUCTION

Increased frequency reuse has been an attractive solution to provide coverage and accommodate more users in cellular networks with precious spectrum resources. In conventional cellular networks based on single-cell processing, the users in a cell are only served by the base station (BS) centered in that cell and a BS is not supposed to assist in enhancing the channel links not belonging to its cell, while the intercell interference (ICI) is dealt with by careful frequency planning and usually treated as noise at the receiver. However, in this single-cell approach, the cell-edge users tend to suffer severe ICI and exhibit much poorer performance than the users in the interior of the cell.

To overcome this, multicell cooperation between BSs has recently emerged as a promising solution, which has been extensively studied in both academic research and standard bodies such as LTE-Advanced/IEEE 802.16m. The idea is that neighboring cells form a cooperating cluster by exchanging information with each other. To be practically useful, only a limited number of cooperating cells should be clustered for the coordinated transmission/reception. In this regard, several solutions including network-centric [1], user-centric methods [2–4], or combinations thereof [5] have been developed recently. The static network-centric method is simple for implementation but users at the cluster edges would still suffer severe out-of-cluster interference. This motivated the development of dynamic user-centric clustering methods [2–4], which allow the cell clusters to be partially overlapping, resulting in a significant increase in scheduling complexity. To tradeoff between performance and complexity, a combination of these two methods was recently proposed [5]. It is realized by using the user-centric method within a set of candidate clusters predetermined via a network-centric means, and is widely adopted by standard bodies to study the multicell cooperation.

Another issue for the cooperative multicell system design is how to optimize the cooperative transmission/reception between the cells. Early work in the area largely focused on the uplink [6, 7], where several BSs jointly decode the signals from the users served by these cells. The problem of downlink multicell processing, however, is much more challenging, especially with multiple antennas at each BS. If a joint transmission strategy is employed, in which both data and channel state information (CSI) are shared between the BSs, the cooperating cluster will essentially collapse into a single-cell multiuser multi-antenna system. Its potential gain has been investigated in [8, 9]. Nevertheless, it is noted that per-BS power constraints should be carefully addressed and an optimal joint transmission requires full phase coherence between the signals from different BSs, which is difficult to realize in practice. Moreover, the required signaling between the BSs is vast, making it challenging for practical applications. Therefore, a distributed solution with limited intercell communication is highly desirable.

More recently, the achievable rate performance of downlink multicell processing in decentralized form [10, 11] and with limited backhaul capacity [12] was analyzed, and showed a large potential. Also, several schemes for distributed multicell processing have been developed in [13–15]. In particular, given the availability of data sharing, [14] addressed the design of distributed transmit beamforming by recasting the downlink beamforming problem into a linear minimum mean-square-error (LMMSE) estimation problem. However, the requirement of information exchange between the BSs is still too high and global convergence is not guaranteed. Later, [15] devised a distributed design using only local CSI, and demonstrated that performance close to the Pareto boundary can be obtained. The main issue for [14, 15] is that both need data sharing between the BSs, which is not likely to be viable due to the limited backhaul capacity.

Motivated by this, [16–20] studied coordinated beamforming techniques without data sharing. In [16], Lagrangian duality theory was applied to design coordinated beamforming with limited intercell information exchange but the solution only applies to the sum-power minimization problem with a signal-to-interference plus noise ratio (SINR) constraint. Also, restricting to limited intercell communication, [17] presented a hierarchical iterative solution to maximize the worst-user rate, but global optimality is only obtained in the case of a total BS power constraint and extensions to per-BS power constraints are nontrivial. It is also possible to devise distributed schemes for coordinated precoding based on the signal-to-generated-interference plus noise ratio (SGINR) [18] and the signal-to-caused-interference ratio (SCIR) [19], with each BS performing optimization only based on local CSI. However, in most cases this strategy is only sum-rate optimal in the high signal-to-noise ratio (SNR) regime with two BSs.

In this paper, our objective is to address the distributed realization of coordinated multicell multiuser downlink beamforming with limited intercell communication. Considering per-BS power constraints, it is known that the optimal rate region achieved by a coordinated multicell system is characterized by the strong Pareto boundary [11], where it is impossible to improve the rate of any user without simultaneously decreasing at least one of the others. With the goal to achieve a fairness based rate optimality, we target at the particular point on the strong Pareto boundary that has the maximum worst-case rate, i.e., the Pareto optimality with max-min fairness. Different from the conventional max-min criterion which is only weakly Pareto optimal[1] and usually produces identical user rates, this objective could achieve a good tradeoff between the sum-rate and user fairness. By recasting the original downlink optimization problem into an approximate dual virtual uplink problem, we will propose an iterative algorithm to reach this desired point.

The main contributions of this paper are as follows:

- We first devise a centralized solution to achieve the Pareto boundary with the max-min rate fairness, which is realized in two steps: (1) a weighted SINR balancing optimization and (2) a Pareto improvement. We prove that our method is able to reach the point of the Pareto boundary on which the worst rate among users is not less than any other points, especially for the case of 2 cooperating BSs.
- Then, we propose a distributed algorithm to approach the performance of the centralized scheme, in which only a small number of positive scalars are exchanged between the BSs. The distributed solution is realized by exploiting an approximate uplink-downlink SINR duality subject to per-BS power constraints. The idea is to iteratively perform the weighted SINR balancing optimization in a distributed fashion by converting it into a virtual uplink problem. Global convergence of the proposed iterative algorithm is proved. Our simulation results also show that the proposed distributed solution obtains a worst-user rate performance close to the optimal centralized scheme while has a better sum-rate performance; and comparable sum-rate performance as the Nash bargaining solution (NBS) but with better max-min fairness.

[1] A point is called weakly Pareto optimal if it is impossible to strictly improve the rates of all the users. If not explicitly stated, the Pareto optimality used in this paper refers to strong Pareto optimality.

This remainder of this paper is organized as follows. Section II describes the system model for coordinated beamforming. Section III presents a two-step centralized multicell downlink beamforming scheme to reach the Pareto boundary with max-min fairness. Section IV analyzes the uplink-downlink SINR duality with per-BS power constraints and proposes a distributed solution. Simulation results are provided in Section V and we conclude the paper in Section VI.

## II. SYSTEM MODEL

Consider a multicell downlink system where cells are allowed to perform coordinated transmission. As shown in LTE-Advanced standardization work for coordinated multi-point transmission/reception (CoMP), it is efficient to dynamically determine the coordinating cell cluster in a user-centric way [21–23]. Each user reports a suggested CoMP cell set based on the reference signal receive power (RSRP) measurement from each cell, where a non-serving cell is included in the reported CoMP set only if its corresponding RSRP is within the given offset of the RSRP of the serving cell, represented by [21]

$$\text{Cell } i \in \text{reporting CoMP set, if and only if} \\ \text{RSRP}_i \geq \text{RSRP}_0 - \alpha_{th}, \quad (1)$$

where $\text{RSRP}_0$ denotes the RSRP of the serving cell, and $\text{RSRP}_i$ denotes the RSRP of the cooperating cell reported by the user, and $\alpha_{th}$ denotes a given threshold whose value is usually set to a small portion of $\text{RSRP}_0$. Then, CoMP user scheduling can be performed and CoMP set will be determined correspondingly. In reality, it is possible that different users report overlapping CoMP sets [21–23], in this case iterative scheduling might be used to achieve a desirable performance. If $\alpha_{th}$ is small, it is understood that the scheduled CoMP users see significant interference from the cooperating cells, where a coordinating transmission is highly desirable. Here we focus on the design of coordinated multicell beamforming within a single cell cluster. Consider a coordinated cell cluster composed of $K$ users and $B$ BSs. Each BS is equipped with $M$ transmit antennas and each user is equipped with a single antenna. Assume that each BS has at least one served user, i.e., $K \geq B$. The transmit signal from each BS is only intended for its own users, meaning that no user-data sharing between the BSs is needed. However, the transmit beamformers from different BSs need to be designed jointly to suppress the ICI.

Let $\mathcal{K}_b$ denote the user set served by BS $b$, and $s_k$ denote the information symbol intended for user $k$, with $\mathbb{E}\{|s_k|^2\} = 1$, where $\mathbb{E}\{\cdot\}$ denotes the expectation operator. Let $b_k$ be the index of the BS serving user $k$, $\mathbf{f}_k \in \mathbb{C}^{M \times 1}$ be the unit-norm beamforming vector (i.e., $\|\mathbf{f}_k\| = 1 \; \forall k$) for user $k$ and $p_k$ be the transmit power for user $k$. Each BS is restrained by a power constraint $P_b$, i.e., $\sum_{k \in \mathcal{K}_b} p_k \leq P_{\text{BS}}$. Assuming $k \in \mathcal{K}_b$, the received signal of user

$k$ is given by

$$y_k = \sqrt{p_k}\mathbf{h}_{kb}^\dagger \mathbf{f}_k s_k + \mathbf{h}_{kb}^\dagger \sum_{\substack{i \in \mathcal{K}_b \\ i \neq k}} \sqrt{p_i}\mathbf{f}_i s_i \\ + \sum_{\bar{b} \neq b} \mathbf{h}_{k\bar{b}}^\dagger \sum_{j \in \mathcal{K}_{\bar{b}}} \sqrt{p_j}\mathbf{f}_j s_j + n_k, \quad (2)$$

where $\dagger$ denotes Hermitian transpose, $n_k$ is the zero-mean complex white Gaussian noise with variance $\sigma_k^2$, and $\mathbf{h}_{kb}$ denotes the frequency-flat fading channel vector from BS $b$ to user $k$. Define $\mathbf{p} \triangleq [p_1,\ldots,p_K]^T$ with the superscript $T$ denoting the transposition, $\boldsymbol{\mathcal{H}}_{ki} \triangleq \frac{\mathbf{h}_{ki}\mathbf{h}_{ki}^\dagger}{\sigma_k^2}$, and $\mathbf{W} \triangleq \mathrm{diag}(\mathbf{f}_1,\mathbf{f}_2,\ldots,\mathbf{f}_K)$. The rate achieved by user $k$ is then written as

$$R_k = \log_2\left(1 + \mathrm{SINR}_k^{\mathsf{DL}}(\mathbf{W},\mathbf{p})\right) \quad (3)$$

where

$$\mathrm{SINR}_k^{\mathsf{DL}}(\mathbf{W},\mathbf{p}) = \frac{p_k \mathbf{f}_k^\dagger \boldsymbol{\mathcal{H}}_{kb} \mathbf{f}_k}{\sum_{\substack{j=1 \\ j \neq k}}^K p_j \mathbf{f}_j^\dagger \boldsymbol{\mathcal{H}}_{k,\Pi(j)} \mathbf{f}_j + 1}, \quad \text{for } k \in \mathcal{K}_b, \quad (4)$$

where $\Pi(j)$ denotes the index of the BS serving user $j$, Note that although the above model appears to assume perfect symbol-to-symbol synchronization, this assumption can be removed in orthogonal frequency-division multiplexing (OFDM) systems when the SINR or some utility function monotonically increasing with SINR is used as the performance metric because the SINR criterion only involves the power of the interference.

The problem considered in this paper, is the joint optimization of the BS beamforming vectors $\{\mathbf{f}_k\}_{\forall k}$ and the power allocation $\{p_k\}_{\forall k}$. A TDD multicell system is first investigated, where each BS (say $b$) can directly estimate the CSI of its own users, i.e., $\{\mathbf{h}_{kb}\}_{k \in \mathcal{K}_b}$, without information exchange between cells, by exploiting downlink-uplink reciprocity. It is also assumed that each BS can estimate from the reverse link the crosstalk channels to other users in the same cluster. This is justifiable because usually the threshold $\alpha_{th}$ is small and therefore the channel from the BS to the users in other cells and that from the BS to its own users are of similar order of magnitude. To do so, a coordinated training phase for the cell cluster might be required, in which the training signals from different users are designed to be orthogonal in order to avoid mutual interference. In summary, $\{\mathbf{h}_{kb}\}_{k=1}^K$ is referred to as local CSI at a TDD BS $b$ and will be exploited in the subsequent design. Note that our focus is not limited to TDD multi-cell systems. Our developing schemes can be easily generalized to frequency division-duplex (FDD) systems via CSI feedback from user terminals, to be discussed later.

## III. Centralized Multicell Beamforming

Various criteria have been used to design multi-cell beamforming schemes, including the sum-rate maximization, the worst-user-rate maximization and the sum-power minimization with SINR constraints. In contrast to these objectives, our design focuses on the tradeoff between rate efficiency and fairness, as each coordinating cell would expect that a certain cooperation gain is guaranteed. If we employ a game-theoretic model and view users in the multi-cell coordinating systems as players with conflicting objectives, it can be shown that the cooperative NBS could achieve a reasonable compromise between efficiency and fairness [10, 24, 25]. However, the algorithms usually require sharing of global information between the BSs and it is hard to realize a fully distributed solution. With the goal to achieve distributed multicell beamforming optimization, we alternatively resort to the max-min fairness based optimization and take the objective as the max-min point on the Pareto boundary of user rates. To gain some insight on our distributed solution, we will first provide a centralized algorithm consisting of two steps. The first step solves a weighted SINR balancing problem in order to reach the max-min fairness with the least sum-power, which is followed by the second step of Pareto improvement targeting at the Pareto optimality. The detailed algorithm is described as follows.

### A. Two-step Optimization

To achieve the max-min fairness, here we take the minimal SINR as the objective function, i.e., to maximize the worst-user SINR. The optimization problem is formulated as

$$\max_{\mathbf{W},\mathbf{p}} \min_k \frac{\mathrm{SINR}_k^{\mathsf{DL}}(\mathbf{W},\mathbf{p})}{\rho_k} \quad \text{s.t.} \quad \sum_{k \in \mathcal{K}_b} p_k \leq P_{\mathrm{BS}} \ \forall b, \quad (5)$$

where $\rho_k$ is a weighting factor to take into account heterogeneous requirements from users; users with larger factors will achieve proportionally higher SINRs. Note that though here we adopt the minimum weighted SINR as the objective function of the max-min optimization, i.e., to maximize the worst-user weighted SINR, this is equivalent to the max-min rate optimization, as the transmission rate is monotonically increasing with the SINR. Problem (5) is difficult to solve due to its non-convexity. Therefore, we first explore its relationship with the following sum-power minimization problem

$$\min_{\mathbf{W},\mathbf{p}} \|\mathbf{p}\|_1 \quad \text{s.t.} \quad \begin{cases} \frac{\mathrm{SINR}_k^{\mathsf{DL}}}{\rho_k} \geq \gamma \ \forall k, \\ \sum_{k \in \mathcal{K}_b} p_k \leq P_{\mathrm{BS}} \ \forall b, \end{cases} \quad (6)$$

in which $\gamma$ is a preset target and $\|\cdot\|_1$ returns the 1-norm of an input vector. It is easily seen that the solution to (6) attains a set of equal weighted SINRs. If we denote the optimal weighted SINR of (5) as $\gamma^*$ and the corresponding beamforming vectors and power vector as $\mathbf{W}^*$ and $\mathbf{p}^*$, respectively, then it can be proved, see [26], that the solution to (6) with a target $\gamma^*$ will produce the same optimal $\mathbf{W}^*$ and $\mathbf{p}^*$. As a result, if (6) is solved, (5) can be simultaneously solved by searching for the highest target $\gamma^*$ for which the solution $\mathbf{p}^*$ of (6) has at least one active power constraint. Furthermore, as shown in [20], (6) can be transformed into a standard second-order cone programming (SOCP) problem and hence can be solved using standard optimization packages, such as CVX [27]. Based on this, a bisection search method can be employed to efficiently search for the optimal weighted SINR target $\gamma^*$, hence giving the solution to (5).

Numerical results show that to achieve the above solution, in many cases only a subset of coordinating BSs need to transmit at full power, while other BSs transmit with less power. Therefore in general it is possible to further improve the sum-rate without losing the max-min optimality. This implies that the above weighted SINR balancing solution is not Pareto optimal. Motivated by this, we next propose a Pareto improvement method in order to reach the Pareto boundary with max-min fairness.

Assuming $M \geq K$, the proposed Pareto improvement is realized by the following beamforming updating algorithm. With a slight notation abuse, let $\{\mathbf{f}_k^*\}_{\forall k}$ and $\mathbf{p}^* = [p_1^*, \ldots, p_K^*]^T$ denote the unit-norm beamforming and power solutions to (5), respectively. For BS $b$, if $\sum_{i \in \mathcal{K}_b} p_i^* < P_{\text{BS}}$ and $p_k^* = \min_{i \in \mathcal{K}_b} p_i^*$, then we propose that user $k$ transmits at full power and its unit-norm beamforming vector is updated as

$$\mathbf{f}_k^{\text{new}} = \frac{\sqrt{p_k^*}\mathbf{f}_k^* + \alpha_k e^{j\theta_k}\mathbf{h}_k^{\text{ZF}}}{\sqrt{P_{\text{BS}}}}, \quad (7)$$

where $\theta_k = \angle(\mathbf{h}_{kb}^\dagger \mathbf{f}_k^*)$, $\alpha_k$ is a positive scalar such that $\|\mathbf{f}_k^{\text{new}}\| = 1$, and $\mathbf{h}_k^{\text{ZF}}$ is the projection of $\mathbf{h}_{kb}$ onto the complement of the column space of $\bar{\mathbf{H}}_k = [\mathbf{h}_{1b} \cdots \mathbf{h}_{k-1,b} \ \mathbf{h}_{k+1,b} \cdots \mathbf{h}_{Kb}]$, given by

$$\mathbf{h}_k^{\text{ZF}} = \left(\mathbf{I} - \bar{\mathbf{H}}_k \left(\bar{\mathbf{H}}_k^\dagger \bar{\mathbf{H}}_k\right)^{-1} \bar{\mathbf{H}}_k^\dagger\right) \mathbf{h}_{kb}. \quad (8)$$

This completes the proposed two-step multicell beamforming scheme.

### B. Pareto Optimality

Here we will discuss the effectiveness of the above proposed scheme, both in terms of max-min fairness and Pareto optimality. In the first step the weighted SINR balancing optimization guarantees the max-min fairness, weighted by the factors $\{\rho_k\}$. We will prove that in the second step the Pareto improvement method yields a rate increase without losing the achieved max-min fairness. For the special case of two BSs and two users, i.e., $B = K = 2$, Pareto optimality of the proposed scheme is guaranteed.

To proceed, we first look at the interference of the signal of user $k$ on other users. The property $\mathbf{h}_{ik}^\dagger \mathbf{h}_k^{\text{ZF}} = 0$ yields that, with the updated beamforming vector $\mathbf{f}_k^{\text{new}}$, the interference power caused on any other user $i \neq k$ is given by

$$P_{\text{BS}} \left|\mathbf{h}_{ib}^\dagger \mathbf{f}_k^{\text{new}}\right|^2 = p_k^* \left|\mathbf{h}_{ib}^\dagger \mathbf{f}_k^*\right|^2. \quad (9)$$

This means that the beamforming update in the second step will not change the interference level imposed on other users, i.e., it will not harm the max-min optimality achieved by the weighted SINR balancing. However, the beamforming update is able to further improve the user's rate. The effective power of user $k$ with the beamforming update can be written as (10), shown at the top of the next page, where equality holds only if the two terms on the right-hand-side have the same phase, i.e., $\theta_k = \angle(\mathbf{h}_{kb}^\dagger \mathbf{f}_k^*)$. That is, the proposed beamforming update method is able to improve the rate as much as possible while keeping the max-min optimality. To further illustrate the efficiency of the proposed two-step optimization algorithm, it is interesting to investigate its Pareto optimality.

As seen from (7), the proposed beamforming update for user $k \in \mathcal{K}_b$ restricts its beamforming update only in the null subspace of $\bar{\mathbf{H}}_k$. Under this subspace restriction, the proposed update method achieves the maximum rate improvement, but there might still be a possibility that an update along some other subspace gives a further rate improvement. Though the proposed algorithm cannot ensure Pareto optimality in the general case, the following theorem shows that Pareto optimality with max-min fairness indeed is achieved in the two-user case.

*Theorem 1:* For the two-BS two-user coordinating system with $M \geq K = B = 2$, the proposed two-step optimization algorithm guarantees the Pareto optimality with max-min fairness.

*Proof:* See Appendix A. ∎

Note that if the proposed weighted SINR balancing optimization suggests that one BS transmits at less power, the proposed beamforming update method would achieve a maximum rate improvement and thus guarantee the Pareto optimality.

## IV. DISTRIBUTED MULTICELL BEAMFORMING

The above solution is by itself centralized as far as implementation is concerned. Here, our aim is to develop a distributed multicell optimization algorithm that reduces the amount of intercell communication. For the step of Pareto improvement, it is seen from (7) that the beamforming vector update for each BS can be performed distributively using local CSI. The main difficulty, however, lies in the distributed implementation of the weighted SINR balancing optimization, which will be addressed by investigating an approximate uplink-downlink duality.

### A. Uplink-Downlink SINR Duality

If we view the cluster of BSs as a super BS (SBS), our system model can be regarded as a virtual multiuser system consisting of $K$ user terminals and a SBS.[2] Previous results in [28, 29] have shown that for each user, say user $k$, the same SINR target can be achieved in both downlink and uplink with the same set of beamforming vectors and the same sum-power. This implies that only if a sum-power constraint for the virtual multiuser system is considered, the existing uplink-downlink duality can be directly applied to solve the downlink optimization problem by converting into a virtual uplink problem, which enables a distributed implementation. When per-BS power constraints are considered, the conventional uplink-downlink duality becomes inapplicable. It is worth mentioning that a modified form of duality proposed in [28] is able to deal with the issue of per-BS power constraints and find the exact optimal solution. However, the subgradient-based algorithm in [28] shows a very slow convergence and this brings difficulties to implementation. As shown in [17], the inter-BS communication required in its implementation involves about $\mathcal{O}(NK)$ positive scalars, where the required number of iterations $N$ is typically about one hundred as shown, thus imposing a heavy burden on the backhaul link. Balancing the pros and cons, here, we propose to use an *approximate* uplink-downlink duality to achieve a distributed multicell beamforming scheme with limited intercell communication.

---

[2]It should be noted that data sharing between the BSs is not permitted in our system model. Therefore, the multiuser precoder of the SBS should be restricted to be in a block-diagonal form expressed as $\text{diag}\{\mathbf{f}_1, \ldots, \mathbf{f}_K\}$.

$$\left|\sqrt{P_{\text{BS}}}\mathbf{h}_{kb}^{\dagger}\mathbf{f}_{k}^{\text{new}}\right|^{2} = \left|\sqrt{p_{k}^{*}}\mathbf{h}_{kb}^{\dagger}\mathbf{f}_{k}^{*} + \alpha e^{j\theta_{k}}\mathbf{h}_{kb}^{\dagger}\left(\mathbf{I} - \bar{\mathbf{H}}_{k}\left(\bar{\mathbf{H}}_{k}^{\dagger}\bar{\mathbf{H}}_{k}\right)^{-1}\bar{\mathbf{H}}_{k}^{\dagger}\right)\mathbf{h}_{kb}\right|^{2}$$
$$\leq \left|\left|\sqrt{p_{k}^{*}}\mathbf{h}_{kb}^{\dagger}\mathbf{f}_{k}^{*}\right| + \left|\alpha\mathbf{h}_{kb}^{\dagger}\left(\mathbf{I} - \bar{\mathbf{H}}_{k}\left(\bar{\mathbf{H}}_{k}^{\dagger}\bar{\mathbf{H}}_{k}\right)^{-1}\bar{\mathbf{H}}_{k}^{\dagger}\right)\mathbf{h}_{kb}\right|\right|^{2}, \quad (10)$$

With the goal to solve the max-min optimization, we develop a virtual uplink problem which is approximately dual to (6) and facilitates a limited-intercell-communication distributed realization, given by

$$\min_{\mathbf{W},\mathbf{q}} \|\mathbf{q}\|_1 \quad \text{s.t.} \quad \begin{cases} \dfrac{\text{SINR}_k^{\text{UL}}}{\rho_k} \geq \gamma \ \forall k, \\ \sum_{k \in \mathcal{K}_b} q_k \leq P_{\text{BS}} \ \forall b, \end{cases} \quad (11)$$

in which $\mathbf{q} = [q_1, \ldots, q_K]^T$ denotes the virtual uplink power vector and

$$\text{SINR}_k^{\text{UL}}(\mathbf{f}_k, \mathbf{q}) = \frac{q_k \mathbf{f}_k^{\dagger} \boldsymbol{\mathcal{H}}_{k,\Pi(k)} \mathbf{f}_k}{\mathbf{f}_k^{\dagger} \left( \sum_{\substack{i=1 \\ i \neq k}}^{K} q_i \boldsymbol{\mathcal{H}}_{i,\Pi(k)} + \mathbf{I} \right) \mathbf{f}_k} \quad (12)$$

denotes the uplink SINR of user $k$. Next we will clarify the approximate duality between (11) and (6).

Denote the solution to (11) as $\mathbf{W}^*$ and $\mathbf{q}^* = [q_1^*, \ldots, q_K^*]^T$, where $\mathbf{q}^*$ satisfies the per-cell power constraints $\sum_{k \in \mathcal{K}_b} q_k^* \leq P_{\text{BS}}$. We convert this uplink power solution $\mathbf{q}^*$ to the downlink using the uplink-to-downlink converting method in [29], and denote the downlink power solution as $\tilde{\mathbf{p}}^* = [\tilde{p}_1^*, \ldots, \tilde{p}_K^*]^T$, which satisfies $\|\tilde{\mathbf{p}}^*\|_1 = \|\mathbf{q}^*\|_1$. The conventional uplink-downlink duality theory reveals that the uplink SINR achieved by $\mathbf{W}^*$ and $\mathbf{q}^*$ is equal to the downlink SINR achieved by $\mathbf{W}^*$ and $\tilde{\mathbf{p}}^*$. Note that strictly speaking $\mathbf{W}^*$ and $\tilde{\mathbf{p}}^*$ are not exactly the solution to (6), as $\tilde{\mathbf{p}}^*$ cannot be guaranteed to satisfy the per-BS power constraints. However, it is found that for any achievable $\gamma$, the equality $\max_b \sum_{k \in \mathcal{K}_b} \tilde{p}_k^* = \max_b \sum_{k \in \mathcal{K}_b} q_k^* \leq P_{\text{BS}}$ approximately holds in most cases, as shown in Fig. 3 (which will be mathematically justified later), implying that $\tilde{\mathbf{p}}^*$ approximately satisfies the per-BS power constraints $\sum_{k \in \mathcal{K}_b} \tilde{p}_k^* \leq P_{\text{BS}}$. This property reveals an approximate duality between the uplink problem (11) and the downlink problem (6), i.e., the same SINR can be achieved by the uplink and the downlink with the same set of beamforming vectors and approximately the same individual (per-cell or per-BS) power constraints. Based on this observation, the solution to the downlink problem (6) can be approximately achieved by solving the uplink counterpart (11).

To mathematically justify this approximate duality, we define

$$\mathbf{D}(\mathbf{W}^*) \triangleq \text{diag}\left\{ \frac{\rho_1 \gamma}{(\mathbf{f}_1^*)^{\dagger} \boldsymbol{\mathcal{H}}_{1,\Pi(1)} \mathbf{f}_1^*}, \ldots, \frac{\rho_K \gamma}{(\mathbf{f}_K^*)^{\dagger} \boldsymbol{\mathcal{H}}_{K,\Pi(K)} \mathbf{f}_K^*} \right\}. \quad (13)$$

and a $K \times K$ matrix $\boldsymbol{\Psi}(\mathbf{W}^*)$ with its $(i,j)$-th element given by

$$[\boldsymbol{\Psi}]_{ij} = \begin{cases} (\mathbf{f}_j^*)^{\dagger} \boldsymbol{\mathcal{H}}_{i,\Pi(j)} \mathbf{f}_j^* & \text{for } i \neq j, \\ 0 & \text{for } i = j. \end{cases} \quad (14)$$

It was shown in [29] that the downlink and uplink power vectors can be obtained as

$$\tilde{\mathbf{p}}^* = \boldsymbol{\Lambda}\mathbf{1}, \quad (15)$$
$$\mathbf{q}^* = \boldsymbol{\Lambda}^T\mathbf{1}, \quad (16)$$

where

$$\boldsymbol{\Lambda} = [\mathbf{I} - \mathbf{D}(\mathbf{W}^*)\boldsymbol{\Psi}(\mathbf{W}^*)]^{-1}\mathbf{D}(\mathbf{W}^*) \quad (17)$$
$$= \left[\mathbf{D}^{-1}(\mathbf{W}^*) - \boldsymbol{\Psi}(\mathbf{W}^*)\right]^{-1},$$

and $\mathbf{1}$ is the all-one vector. It is difficult to derive a rigorous condition for the uplink-downlink duality with individual power constraints, i.e., $\max_b \sum_{k \in \mathcal{K}_b} \tilde{p}_k^* = \max_b \sum_{k \in \mathcal{K}_b} q_k^* \leq P_{\text{BS}}$. As an alternative, our analysis resorts to examining how far from being diagonal $\mathbf{A} \triangleq \mathbf{D}^{-1}(\mathbf{W}^*) - \boldsymbol{\Psi}(\mathbf{W}^*)$ is, as the duality holds exactly when it is strictly diagonal. That is, we will examine the duality between (6) and (11) by seeing if $\mathbf{A}$ is diagonal. We will show that (11) is diagonal-dominant or quasi-diagonal in most cases, i.e., the duality approximately holds.

To proceed, we define the following metric:

$$\eta_k = \frac{(\mathbf{f}_k^*)^{\dagger} \boldsymbol{\mathcal{H}}_{k,\Pi(k)} \mathbf{f}_k^*}{\rho_k \gamma \sum_{\substack{i=1 \\ i \neq k}}^{K} (\mathbf{f}_k^*)^{\dagger} \boldsymbol{\mathcal{H}}_{i,\Pi(k)} \mathbf{f}_k^*} \quad (18)$$

which is the ratio of the magnitude of diagonal entry in the $k$-th column to the sum of the magnitudes of other entries in this column. A matrix $\mathbf{A}$ is said to be strictly diagonal-dominant if $\eta_k > 1 \ \forall k$ [30]. Recalling that $\{\mathbf{f}_k^*\}$ is the solution to (11), we therefore have

$$\rho_k \gamma \leq \frac{q_k^* (\mathbf{f}_k^*)^{\dagger} \boldsymbol{\mathcal{H}}_{k,\Pi(k)} \mathbf{f}_k^*}{(\mathbf{f}_k^*)^{\dagger} \left( \sum_{\substack{i=1 \\ i \neq k}}^{K} q_i^* \boldsymbol{\mathcal{H}}_{i,\Pi(k)} + \mathbf{I} \right) \mathbf{f}_k^*}. \quad (19)$$

Based on the above, $\eta_k$ can be rewritten as

$$\eta_k \geq \frac{\sum_{\substack{i=1 \\ i \neq k}}^{K} q_i^* (\mathbf{f}_k^*)^{\dagger} \boldsymbol{\mathcal{H}}_{i,\Pi(k)} \mathbf{f}_k^* + 1}{\sum_{\substack{i=1 \\ i \neq k}}^{K} q_k^* (\mathbf{f}_k^*)^{\dagger} \boldsymbol{\mathcal{H}}_{i,\Pi(k)} \mathbf{f}_k^*}. \quad (20)$$

Note that typical user-centric cell clustering strategies [21–23] result in that user-BS links have comparable channel strengths, resulting in that $\{q_i^*\}_{i=1}^{K}$ will have the same order of magnitudes to yield $\eta_k > 1 \ \forall k$.

It is also useful to examine the asymptotic behavior of $\mathbf{A}$. When the received SNR becomes extremely small, i.e., $q_k^* \boldsymbol{\mathcal{H}}_{i,\Pi(k)} = \frac{q_k^* \mathbf{h}_{i,\Pi(k)} \mathbf{h}_{i,\Pi(k)}^{\dagger}}{\sigma_i^2} \approx \mathbf{0}$, it is easy to see that $\eta_k \gg 1$. This implies that $\mathbf{A}$ and also $\boldsymbol{\Lambda}$ are quasi-diagonal. Similarly, when the received SNR becomes extremely large, the inter-user interference becomes dominant over the noise. Thus, the main purpose of the beamforming optimization in solving (11) now lies in mitigating the inter-user interference, leading to $(\mathbf{f}_k^*)^{\dagger} \boldsymbol{\mathcal{H}}_{i,\Pi(k)} \mathbf{f}_k^* \approx 0 \ \forall i \neq k$. This yields again $\eta_k \gg 1$.

## B. Distributed Implementation

To achieve the weighted SINR balancing problem (5) in a distributed way, here we propose an iterative algorithm to achieve the solution to the approximate virtual uplink problem, and then convert it to the downlink. Based on the derived approximate duality, the virtual uplink problem corresponding to (5) is formulated as

$$\max_{\mathbf{W},\mathbf{q}} \min_{k} \frac{\text{SINR}_k^{\text{UL}}(\mathbf{W},\mathbf{q})}{\rho_k} \quad \text{s.t.} \quad \sum_{k \in \mathcal{K}_b} q_k \le P_{\text{BS}} \; \forall b. \quad (21)$$

Using a similar method as the one in solving (6) in Section III-A, (21) can be solved by iteratively solving the sum-power minimization (11) and searching for a maximum achievable SINR target. As shown in [17], such a strategy permits a distributed solution using a standard power control loop [31]. However, its implementation requires a two-layer iteration and suffers from slow convergence. To speed up the convergence, we propose a single-layer iterative algorithm to reach a distributed solution.

**Proposed Iterative Algorithm**

0) Use the superscript $(n)$ to denote the corresponding parameters in the $n$th iteration.
1) Initialize $\mathbf{q}^{(0)}$ with $q_k^{(0)}$ drawn from a uniformly distributed random variable between $(0, P_{\text{BS}}]$ or a simple all-zero vector, $\gamma^{(0)} = 10^{-3}$ and $n = 1$.
2) Set $g_k^{(n)} = \gamma^{(n-1)} I_k(\mathbf{q}^{(n-1)})$ for all $k$, where the interference function $I_k(\mathbf{q}^{(n)})$ is defined as

$$I_k(\mathbf{q}^{(n)}) \triangleq \frac{\sum_{i=1, i \ne k}^{K} q_i^{(n)} (\mathbf{f}_k^{(n)})^{\dagger} \mathcal{H}_{i, \Pi(k)} \mathbf{f}_k^{(n)} + 1}{(\mathbf{f}_k^{(n)})^{\dagger} \mathcal{H}_{k, \Pi(k)} \mathbf{f}_k^{(n)}} \quad (22)$$

in which

$$\mathbf{f}_k^{(n)} = \arg\max_{\mathbf{f}_k} \frac{q_k^{(n)} \mathbf{f}_k^{\dagger} \mathcal{H}_{k, \Pi(k)} \mathbf{f}_k}{\mathbf{f}_k^{\dagger} \left( \sum_{i=1, i \ne k}^{K} q_i^{(n)} \mathcal{H}_{i, \Pi(k)} + \mathbf{I} \right) \mathbf{f}_k}. \quad (23)$$

3) Find $\alpha = \min_b \frac{P_{\text{BS}}}{\sum_{k \in \mathcal{K}_b} g_k^{(n)}}$.
4) Update $q_k^{(n)} = \alpha g_k^{(n)}$ for all $k$.
5) Find the minimum SINR and set $\gamma^{(n)} = \min_k \text{SINR}_k^{\text{UL}}(\mathbf{f}_k^{(n)}, \mathbf{q}^{(n)})$.
6) Stop if the convergence condition defined by $\|\mathbf{q}^{(n)} - \mathbf{q}^{(n-1)}\| < \varepsilon$ (a positive scalar close to zero) is satisfied. Otherwise, $n = n + 1$ and go back to Step 2.

*Theorem 2:* The proposed iterative algorithm with any arbitrary initialization always converges to the global optimum of (21).

*Proof:* We define $\mathcal{I}(\gamma, \mathbf{q}) \triangleq [\gamma I_1(\mathbf{q}), \ldots, \gamma I_K(\mathbf{q})]^T$, which can be recognized as a standard interference function [31] for the SINR target $\gamma$. We also find it useful to first review the key properties of the fixed-point iteration algorithm. Assuming a feasible initial point $\mathbf{q}^{(0)}$ to achieve $\gamma$, the fixed-point iteration $\mathbf{q}^{(n)} = \mathcal{I}(\gamma, \mathbf{q}^{(n-1)})$ has the following properties [31]:

P1: $\mathbf{q}^{(n)}$ is component-wise monotonically decreasing;
P2: $\mathbf{q}^{(n)}$ converges to the unique optimal solution;
P3: $\mathbf{q}^{(n)}$ for all $n$ are feasible solutions.

Our proof is divided into two parts. We first prove its convergence to a fixed point, and then prove that this fixed point is optimal. We start with Step 5 in the $n$th iteration. Obviously, $\mathbf{q}^{(n)}$ satisfies the individual power constraints and achieves $\gamma^{(n)}$. At Step 2 in the $(n+1)$th iteration, $\mathbf{g}^{(n+1)} = \mathcal{I}(\gamma^{(n)}, \mathbf{q}^{(n)})$ is obtained via a standard mapping. Then, according to P1, we have

$$\mathbf{g}^{(n+1)} \le \mathbf{q}^{(n)}. \quad (24)$$

As a result, at the following Step 3, $\alpha \ge 1$. Based on the SINR expression in (12), this yields

$$\begin{aligned}\text{SINR}_k^{\text{UL}}(\mathbf{f}_k^{(n+1)}, \mathbf{q}^{(n+1)}) &= \text{SINR}_k^{\text{UL}}(\mathbf{f}_k^{(n+1)}, \alpha \mathbf{g}^{(n+1)}) \\ &\ge \text{SINR}_k^{\text{UL}}(\mathbf{f}_k^{(n+1)}, \mathbf{g}^{(n+1)})\end{aligned} \quad (25)$$

According to Step 2 and the definition of $\gamma^{(n)}$, we have

$$\min_k \text{SINR}_k^{\text{UL}}(\mathbf{f}_k^{(n+1)}, \mathbf{g}^{(n+1)}) \ge \gamma^{(n)}. \quad (26)$$

Therefore, $\gamma^{(n+1)} \ge \gamma^{(n)}$, i.e., the minimum SINR is monotonically increasing and the algorithm converges to a fixed point $\{\gamma^{(\infty)}, \mathbf{q}^{(\infty)}\}$, in which at least one individual power constraint should be active. Next, we proceed to prove that the fixed point is also optimal by contradiction.

Assume that the optimal solution is $\{\gamma^{\text{opt}}, \mathbf{q}^{\text{opt}}\}$ and $\gamma^{(\infty)} < \gamma^{\text{opt}}$. We see that $\mathbf{q}^{(\infty)}$ satisfies that at least one individual power constraint should be active and

$$q_k^{(\infty)} = \gamma^{(\infty)} I_k(\mathbf{q}^{(\infty)}) \; \forall k. \quad (27)$$

As a consequence, $\mathbf{q}^{(\infty)}$ is the unique solution to (11) with the SINR constraint $\gamma^{(\infty)}$ according to P2. Now we take $\mathbf{q}^{\text{opt}}$, which is also a feasible solution as the initial vector for the fixed-point algorithm to solve (11) and the converged power vector is $\tilde{\mathbf{q}}$. Then, according to P1, we have $\tilde{\mathbf{q}} < \mathbf{q}^{\text{opt}}$ which is expected to be the same as $\mathbf{q}^{(\infty)}$. Since $\mathbf{q}^{\text{opt}}$ should satisfy the individual power constraints, this indicates that with $\tilde{\mathbf{q}}$, no individual power constraint is active and therefore contradicts that $\mathbf{q}^{(\infty)}$ is the unique optimal solution to (11). ∎

After we obtain the solution of (21), say $\{\bar{\gamma}, \bar{\mathbf{q}}, \bar{\mathbf{W}}\}$, the solution to the downlink problem can be achieved using a standard power control loop in a distributed manner. However, in order to fully utilize the backhaul link between the cooperating BSs, here we choose to calculate the downlink power vector based on the following linear system of equations [20, (19)–(21)]

$$\bar{\mathbf{p}} = \left[ \mathbf{D}^{-1}(\bar{\mathbf{W}}) - \boldsymbol{\Psi}(\bar{\mathbf{W}}) \right]^{-1} \mathbf{1}. \quad (28)$$

It is possible that some $\bar{p}_k$ obtained from above could violate the individual BS power constraints, due to the fact that the duality with individual power constraints only holds approximately but not exactly. In that case, the power of the specific BS is scaled down to fulfill the power constraints, which would possibly result in some performance loss in terms of the max-min fairness. Our simulation results in Section V will demonstrate that the percentage of the power constrain violation is marginal and its resulting worst-user rate performance loss is insignificant. On the other hand, it is worth mentioning that the possible scaling down of the power solution to the max-min optimization problem would provide more room for the sum-rate improvement in the second step. Simulation results show that our distributed scheme usually achieves a better sum-rate performance though suffering a performance loss in terms of the worst-user rate, compared to the centralized algorithm.

## C. Required Inter-BS Communication

In this subsection, we analyze the required inter-BS communication in the proposed distributed algorithm. As has been noted before, the Pareto improvement can be separately implemented at each BS using local CSI $\{\mathbf{h}_{kb}\}_{k=1}^{K}$, without the need of inter-BS communication. The information exchange between the BSs is only required in the max-min optimization in which the proposed iterative algorithm is first implemented and its solution is then converted to the downlink.

Assume that there are $B$ coordinating BSs simultaneously serving $K$ users, each of which (say BS $b$) has $|\mathcal{K}_b|$ users. By examining the steps of the proposed iterative algorithm, we can see that the results will not be changed if we use $g_k^{(n)} = I_k(\mathbf{q}^{(n-1)})$ in Step 2 instead of $g_k^{(n)} = \gamma^{(n-1)} I_k(\mathbf{q}^{(n-1)})$, so $\gamma^{(n)}$ is not needed in actual calculation. As a consequence, the implementation only requires that the power vector $\mathbf{q}^{(n)}$ at each iteration be shared between the BSs. That is, each iteration requires the exchange of $\sum_{b=1}^{B}(B-1)|\mathcal{K}_b| = K(B-1)$ positive scalars between the BSs[3]. Suppose that the algorithm converges within $(N+1)$ iterations[4], the inter-BS communication involves $NK(B-1)$ positive scalars. Then we look at the computation of the downlink power vector $\bar{\mathbf{p}}$ from the converged result of the iterative algorithm, it can be seen from (28) that each BS can distributively calculate its preferred power by sharing matrices $\mathbf{D}$ and $\boldsymbol{\Psi}$. As shown in the definitions of $\mathbf{D}$ and $\boldsymbol{\Psi}$, it needs $K(B-1)\sum_{b=1}^{B}|\mathcal{K}_b| = K^2(B-1)$ positive scalars of inter-BS communication.

To highlight the light inter-BS communication requirement of the proposed method, we compare it with a simple strategy where full CSI is shared between the BSs and each BS individually performs beamforming optimization using global CSI information. Recognizing that exchange of one $M$-dimension channel vector requires about $4M$ positive scalars, the latter strategy would take about $4MKB(B-1)$ positive scalars of inter-BS communication, which increases linearly with the number of antennas at the BSs. In comparison, the ratio of the inter-BS communication of the proposed method to that of full CSI exchange is calculated as $\frac{N+K}{4MB}$. Our simulations show that the convergence speed of the proposed method is not significantly affected by the number of BS antennas, due to the fact that the number of the dual variables in our problem formulation only depends on the number of coordinating BSs and users but not the number of antennas. This reveals that our proposed method has much reduced requirement for inter-BS communication, especially when the number of transmit antennas is large.

## D. Practical Issues

Note that the proposed distributed scheme assumes that each BS $b$ knows its local CSI $\{\mathbf{h}_{kb}\}_{\forall k}$ perfectly. This is hard to realize in practical systems. It is more reasonable to assume that only imperfect CSI is available at the BSs. In particular, if the bandwidth of the backhaul link is limited, the implementation of the proposed iterative optimization will suffer a fairly large delay because of the parameter exchange between the BSs. Thus, the CSI could be outdated by the time the iterations are completed, resulting in some performance loss. This problem becomes severe in high-mobility scenarios where the channels vary rapidly. Even in low-mobility scenarios, it is still possible that CSI varies while the algorithm is running. Moreover, a certain level of channel estimation error is inevitable in practice. This motivates us to evaluate the robustness of the proposed scheme against imperfect CSI.

Imperfect CSI is commonly modeled either stochastically or deterministically. In the former case one assumes that the uncertainty region of the CSI errors is unbounded and distributed according to some known distribution [32–34], while in the latter case the uncertainty region of the CSI perturbations is assumed to be bounded [35, 36]. Here we will adopt the bounded imperfect CSI model, as we can evaluate the performance of our proposed scheme under a restricted iteration delay. This can be realized by fixing the number of iterations, such that the backhaul signaling overhead is restrained and the resulting delay is also limited. This kind of evaluation is of particular importance for practical applications. By decomposing the channel vector into a large-scale factor and a small-scale fading part, the erroneous channel can be expressed as

$$\hat{\mathbf{h}}_{kb} = \gamma_{kb}\hat{\mathbf{h}}_{kb}^{(\mathrm{s})} = \gamma_{kb}\left(\mathbf{h}_{kb}^{(\mathrm{s})} + \tilde{\mathbf{h}}_{kb}^{(\mathrm{s})}\right), \quad (29)$$

where $\gamma_{kb}$ denotes the large-scale channel component and is assumed to be exactly estimated by the BS, $\mathbf{h}_{kb}^{(\mathrm{s})}$ denotes the actual small-scale fading part of the channel, $\hat{\mathbf{h}}_{kb}^{(\mathrm{s})}$ denotes the nominal small-scale fading channel known by the BS, and $\tilde{\mathbf{h}}_{kb}^{(\mathrm{s})}$ denotes the channel uncertainty. As is common in the literature [35], we assume that the channel uncertainty is confined within an origin-centered hyper-spherical region of radius $\sigma_{kb}$, i.e., $\|\tilde{\mathbf{h}}_{kb}^{(\mathrm{s})}\|_2 \leq \sigma_{kb}$.

Simulation results in Fig.8 will show that for a CSI uncertainty with $\sigma_{kb} = 0.1$, the proposed distributed solution for $M = 4$ only suffers from marginal performance loss. It is also worth noting that, the strategy of robust design in [35] can be readily extended to our scheme. To do so, we only need to define $\mathrm{sinr}_k^{\mathrm{ul}} \triangleq \min_{\{\|\tilde{\mathbf{h}}_{kb}^{(\mathrm{s})}\|_2 \leq \sigma_{kb}\}} \mathrm{SINR}_k^{\mathrm{UL}}$ as the worst-case SINR over the uncertainty regions, and use it to substitute $\mathrm{SINR}_k^{\mathrm{UL}}$ in the objective function. Several possible bounds of $\mathrm{sinr}_k^{\mathrm{ul}}$, e.g., [35, (13)], can be used in our algorithm.

Finally, we will discuss the implementation of our scheme in FDD multicell systems. Different from TDD implementation, in FDD mode the CSI can only be obtained through feedback from users. It is noted that normally each user terminal, say $k$, can only estimate its channels from all the coordinating BSs, i.e., $\{\mathbf{h}_{kb}\}_{\forall b}$, and feed back these CSI to its serving BS. Therefore, in order to generalize our scheme to FDD systems, extra channel information exchanging between the BS will be needed such that each BS $b$ can have CSI $\{\mathbf{h}_{kb}\}_{\forall k}$ and then perform iterative beamforming optimization.

## V. SIMULATIONS RESULTS

In this section, computer simulations are provided to evaluate the performance of the proposed schemes. The user-centric cell clustering strategy described in Section II is used in our simulations. In particular, our user scheduling only considers the cell-edge users reporting the same 2-BS or 3-BS CoMP set with $\alpha_{th} = 0.4$ RSRP$_0$. More specifically, we assume that the inter-BS distance is 1km and the users have at least 350m distance from their serving BS. For each time slot, each BS uses a round-robin strategy to schedule a single user from its serving user pool. We consider flat fading channels, where the small-scale fading $\mathbf{h}_{kb}^{(s)}$ is assumed to be zero-mean Gaussian distributed with covariance $\mathbf{I}$, i.e., $\mathbf{h}_{kb}^{(s)} \sim \mathcal{CN}(\mathbf{0}, \mathbf{I})$, the large-scale path loss $\gamma_{kb}$ is given by

$$\gamma_{kb} = \frac{\beta \mu_{kb}}{d_{kb}^l} \tag{30}$$

in which $\beta = 10^{-3.45}$ is a scaling factor, and $l = 3.8$ denotes the path loss exponent, $d_{kb}$ is the distance between user $k$ and BS $b$. In dB, this gives $10 \log_{10}(\gamma_{kb}) = -38 \log_{10}(d_{kb}) - 34.5 + \mu_{kb}$, where the shadow fading follows a normal distribution $\mu_{kb} \sim \mathcal{N}(0, 8 \text{ dB})$. Throughout the simulations, we set $\rho_k = 1$ and assume that the coordinating BSs have the same power constraint, and the scheduled users have the same noise figure. The user SNR is defined as the ratio of the average signal power from its serving BS to the noise power. For simplicity, we assume the scheduled users have the same path-loss to their serving BSs and thus have the same SNR.

---
[3]This is evaluated based on the signaling overhead of backhaul links. If one BS needs to transmit one positive scalar to other $B-1$ BSs, it would involve an exchange of $B-1$ positive scalars.

[4]Our simulation results show that typically around 5 iterations are sufficient to achieve a desired accuracy.

For comparison, several existing multicell beamforming schemes based on different criteria are also simulated, including the SGINR [18, 19] scheme, the sum-rate optimal scheme and the NBS. The optimal sum-rate is realized by numerical search over the achievable rate set generated using the method in [11, Corollary 2]. The NBS is achieved by solving the optimization problem as follows [10, 24, 25]

$$\max_{\mathbf{W}, \mathbf{p}} \prod_k (R_k - R_k^0) \quad \text{s.t.} \quad \begin{cases} R_k \geq R_k^{(0)} \; \forall k, \\ \sum_{k \in \mathcal{K}_b} p_k \leq P_{\text{BS}} \; \forall b, \end{cases} \tag{31}$$

where $\{R_k^0\}$ denotes the initial agreement point. As is commonly done, we use the Nash Equilibrium (NE) solution $\{R_k^{\text{NE}}\}$ as the initial point. The above NBS problem in general is non-convex, so we employ alternating optimization between $\mathbf{W}$ and $\mathbf{p}$ to find a near-optimal solution. Taking into account the possible limitation of backhaul capacity in practical systems, we also simulate the proposed scheme with a fixed number of iterations, such that the information exchange between the BSs is limited.

The Pareto boundaries of the multicell beamforming system with $(M, K) = (4, 2)$ under two random channel realizations are first simulated and shown in Figs. 1 and 2, using the approach [11, Corollary 2]. It is seen that except the NE point, other schemes all achieve Pareto optimality in these two cases. In particular, as shown in Fig. 1, the rate point achieved by the proposed centralized scheme is located exactly at the intersection point between the Pareto boundary and the line $R_2 = R_1$. In another case as shown in Fig. 2, the Pareto boundary has no intersection point with the line $R_2 = R_1$, implying that the SINR balanced rate point does not operate on the Pareto boundary, but the rate point of our centralized scheme is still on the Pareto boundary and is the point closest to the line $R_2 = R_1$. This means that in these two cases the rate point achieved by the proposed centralized scheme always has its worst rate not less than any other points on the Pareto boundary, guaranteeing a strict max-min fairness. As for the proposed distributed scheme, it is seen that the rate point also is located on the Pareto boundary, between the rate point of the centralized and the sum-rate optimal point, having a behavior similar to the NBS solution. This implies that compared to the centralized scheme, the distributed scheme has a certain performance loss in terms of max-min fairness, due to the fact that the duality is only approximate, although it achieves a better sum-rate performance.

The analysis in Section IV-A has shown that if the condition $\max_b \sum_{k \in \mathcal{K}_b} \tilde{p}_k^* \leq P_{\text{BS}}$ is met, the optimization of the virtual uplink problem can guarantee the efficiency of the resulting downlink solution. Otherwise, if $\max_b \sum_{k \in \mathcal{K}_b} \tilde{p}_k^*$ exceeds the individual power constraint $P_{\text{BS}}$, our distributed scheme will in general suffer from a performance loss in terms of max-min fairness, depending on how much the power constraints are exceeded. For this reason, we define the following excessive power percentage

$$\varphi = \frac{\max_b \sum_{k \in \mathcal{K}_b} \tilde{p}_k^* - P_{\text{BS}}}{P_{\text{BS}}}. \tag{32}$$

Fig. 3 shows the cumulative density function (CDF) of $\varphi$ under several configurations. We see that $\max_k \tilde{p}_k^* \leq P_{\text{BS}}$ holds with a probability of more than 70%, and $\varphi$ has a value below 3% with a probability of more than 90%, showing that the duality holds with high probability.

Fig. 4 illustrates the average rate results achieved by the worst user in the proposed multicell beamforming schemes and the reference schemes for $(M, K) = (4, 3)$, calculated over 5000 random channel realizations. The results reveal that the proposed centralized and distributed schemes both outperform the reference schemes over a wide SNR range. In particular, the centralized scheme achieves a gain around 0.5 bpsk/Hz over the sum-rate optimal scheme, the SGINR scheme and the NBS. In contrast to the centralized scheme, the distributed scheme shows a performance loss around 0.25 bpsk/Hz. Results also show that most performance improvements in our distributed scheme are achieved already within 2 iterations. This means that the efficiency of our distributed scheme can be maintained even with very limited inter-BS communication. Fig. 5 shows the CDF of the worst-user rate results achieved by the above schemes. It is seen that at all outage levels, our centralized scheme exhibits the best worst-user rate performance. Though our distributed scheme suffers a slight worst-user rate loss compared to the centralized scheme, it outperforms the NBS and the sum-rate optimal scheme. Looking at the sum-rate performance, Fig. 6 provides the CDF of the ratio of the sum-rate achieved by the above schemes to the optimal solution. The results demonstrate that the guarantee of strict max-min fairness by our centralized scheme comes at the cost of a large sum-rate performance loss. On the contrary, our two-step optimized distributed scheme achieves a significant sum-rate improvement over the centralized scheme by sacrificing a certain max-min fairness. It is seen that the outage behavior of the sum-rate achieved by our distributed scheme is very close to the NBS. At an outage probability of 10%, more than 80% of the optimal sum-rate is achieved. Combining the results in Fig. 5, in this setup our distributed scheme outperforms the NBS in terms of fairness rate and has a comparable sum-rate performance.

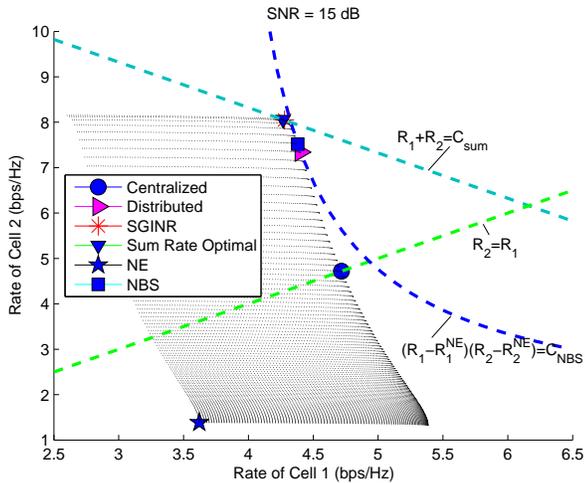

Fig. 1. The rate points of the multicell schemes and the Pareto boundary under a random channel realization, with $(M, K) = (4, 2)$ and SNR=15 dB.

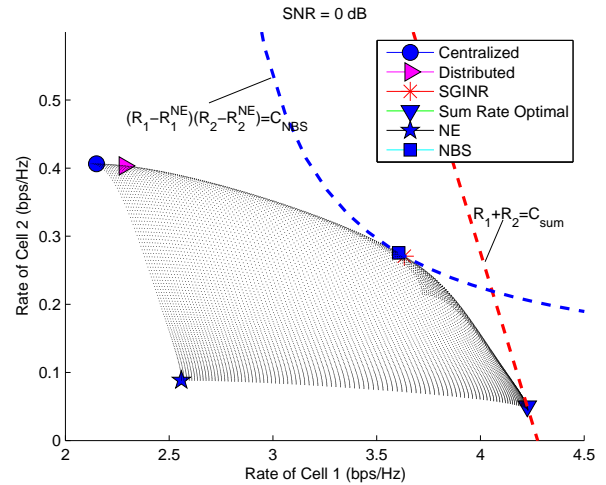

Fig. 2. The rate points of the multicell schemes and the Pareto boundary under a random channel realization, with $(M, K) = (4, 2)$ and SNR=0 dB.

To evaluate the impact of quantized parameter exchange between the BSs on the performance of the proposed distributed scheme, we provide the average worst-user rate performance of the distributed scheme applying a simple linear uniform quantization between 0 and $P_{\text{BS}}$ to the power parameters $\mathbf{q}^{(n)}$. It is seen in Fig. 7 that a 3-bit quantization gives an inappreciable performance loss. Finally, to investigate the robustness of our distributed scheme against the CSI imperfection, we adopt the CSI error model (29) and set $M = 4, \sigma_{kb} = \{0.05, 0.1, 0.2\}$, plotting the average rate of the worst case user assuming that the CSI errors are uniformly distributed in the uncertain region. Simulation results in Fig. 8 show that though CSI errors generally degrade the performance and the loss increases with the SNR, no error floor is observed at SNR = 20 dB in any of the setups. A CSI error level less than $\sigma_{kb} = 0.1$ gives a slight performance loss especially in low and medium SNR regimes. Even with $\sigma_{kb} = 0.2$, the proposed distributed scheme still outperforms the SGINR scheme with accurate CSI over a wide SNR range.

## VI. CONCLUSION

This paper has presented two fair-rate optimal multicell downlink coordinated beamforming schemes. We first devised a two-step centralized optimization scheme, in which the first step achieves a strict max-min rate fairness, while the second step further improves the sum-rate as much as possible without losing the max-min optimality. We proved that this scheme is guaranteed to reach the Pareto boundary with max-min fairness in the case of two BSs each serving a single user. By exploiting the approximate uplink-downlink SINR duality with individual power constraints, a distributed solution, in which only a small number of positive scalars are shared between cells, was proposed. Simulation results showed that our distributed scheme can approach the max-min Pareto optimality, requiring only very limited inter-BS communication, suitable for practical BS cooperation.

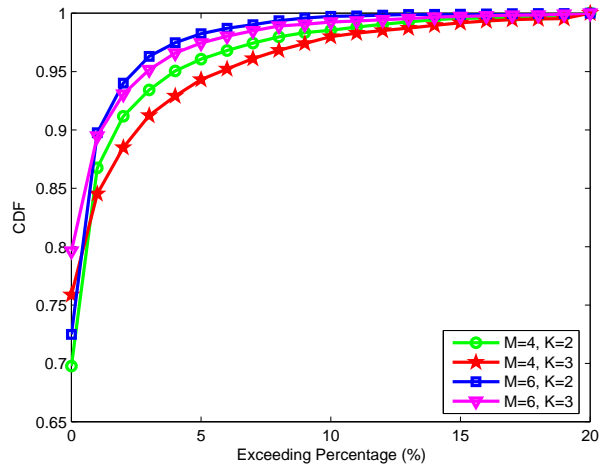

Fig. 3. The CDF of the excessive power percentage for the calculated downlink power vector with SNR=10 dB.

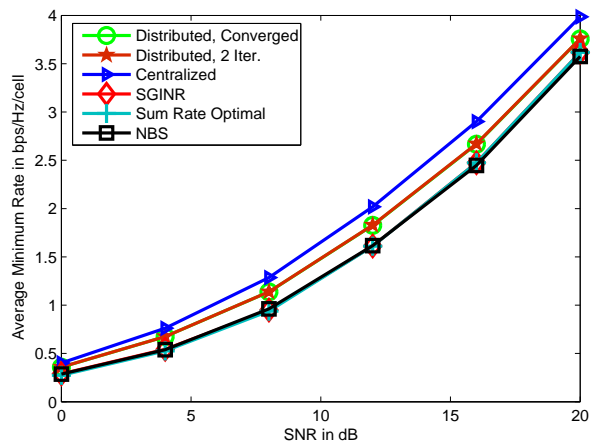

Fig. 4. The average worst-user rates of the proposed multicell schemes and the reference schemes with $K = 3$.

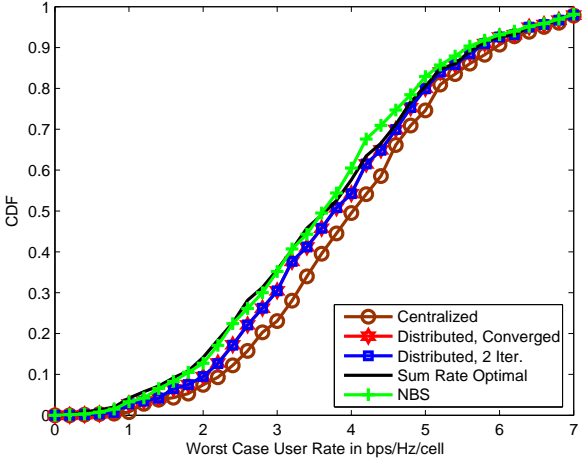

Fig. 5. The CDF of the worst-user rate achieved by the multicell beamforming schemes with $K = 3$ and SNR=20 dB.

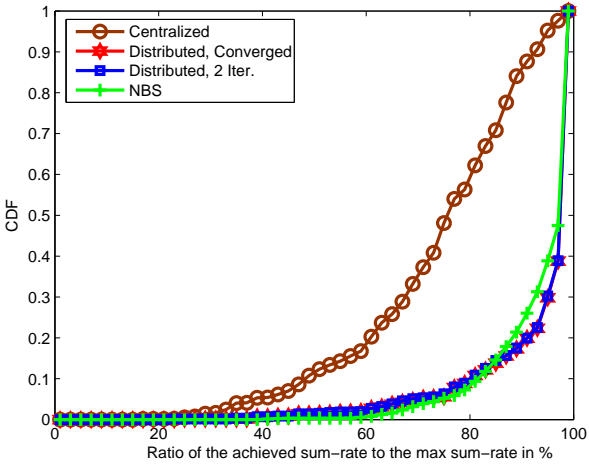

Fig. 6. The CDF of the ratio of the sum-rate of the multicell beamforming schemes to the max sum-rate, with $K = 3$ and SNR=20 dB.

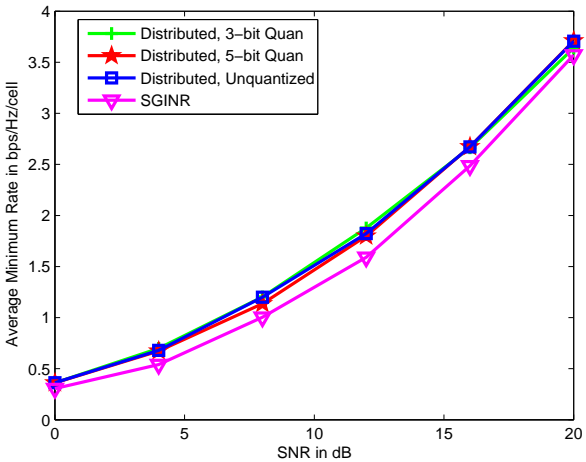

Fig. 7. The average worst-user rate of the proposed distributed multicell scheme employing uniform linear quantization on the exchanged power parameters, and fixing 2 iterations, $K = 3$.

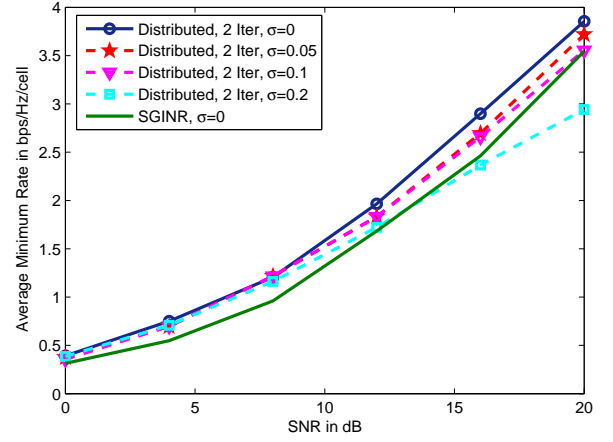

Fig. 8. The average worst-user rate as a function of SNR in the proposed distributed multicell scheme with estimated CSI at each BS, $K = 3$, and 2 iterations.

## APPENDIX A
## PROOF OF THEOREM 1

We consider two separate cases. If the solution to the weighted SINR balancing problem suggests that both BSs use full transmit power, then it is impossible to obtain a Pareto improvement, meaning that the Pareto optimality with max-min fairness has already been achieved. Therefore, we only need to focus on the other case where one BS uses full transmit power while the other one uses less power. Without loss of generality, assume $p_1^* < P_{\text{BS}}$ and $p_2^* = P_{\text{BS}}$. The proof of Theorem 1 starts with the following Lemma.

*Lemma 1:* In the solution to the weighted SINR balancing problem with $M \geq K = B = 2$, suppose $p_1^* < P_{\text{BS}}$ and $p_2^* = P_{\text{BS}}$. If a Pareto improvement over the balanced rate tuple $(r^*, r^*)$ can be achieved in which user 1's rate improves while user 2's rate remains, then the updated beamforming vector for user 1, denoted by $\mathbf{f}_1^{\text{new}}$, must generate no interference to user 2, i.e., $\mathbf{f}_1^{\text{new}\dagger} \boldsymbol{\mathcal{H}}_{21} \mathbf{f}_1^{\text{new}} = 0$.

*Proof:* The proof is by contradiction. First, we let the beamforming vectors and the transmit power of user 2 be fixed by $\mathbf{f}_1^{\text{new}}$, $\mathbf{f}_2^*$ and $P_{\text{BS}}$, respectively. Then, we can express the SINRs (assuming $\rho_k = 1$ for simplicity) of the two users as functions of $p_1$, given by

$$\text{SINR}_1^{\text{DL}}(p_1) = \frac{p_1 \mathbf{f}_1^{\text{new}\dagger} \boldsymbol{\mathcal{H}}_{11} \mathbf{f}_1^{\text{new}}}{P_{\text{BS}} \mathbf{f}_2^{*\dagger} \boldsymbol{\mathcal{H}}_{12} \mathbf{f}_2^* + 1}, \qquad (33)$$

$$\text{SINR}_2^{\text{DL}}(p_1) = \frac{P_{\text{BS}} \mathbf{f}_2^{*\dagger} \boldsymbol{\mathcal{H}}_{22} \mathbf{f}_2^*}{p_1 \mathbf{f}_1^{\text{new}\dagger} \boldsymbol{\mathcal{H}}_{21} \mathbf{f}_1^{\text{new}} + 1}. \qquad (34)$$

Define $\Delta\text{SINR}(p_1) \triangleq \text{SINR}_1^{\text{DL}}(p_1) - \text{SINR}_2^{\text{DL}}(p_1)$. We observe that $\Delta\text{SINR}(p_1)$ is a monotonically increasing and continuous function. The assumption of Pareto improvement thus yields

$$\Delta\text{SINR}(p_1^{\text{new}}) > 0 \qquad (35)$$

but it can be easily observed that $\Delta\text{SINR}(0) < 0$. As a consequence, there must exist some $p_1' < p_1^{\text{new}}$ such that $0 < \Delta\text{SINR}(p_1') < \Delta\text{SINR}(p_1^{\text{new}})$. Now, if we assume that

$\mathbf{f}_1^{\text{new}\dagger} \mathcal{H}_{21} \mathbf{f}_1^{\text{new}} \neq 0$ is true, then this implies that if BS 1 chooses to use transmit power $p_1'$, the resulting SINRs of both users will be greater than $\gamma^*$, which contradicts the result that $\gamma^*$ is the solution to the max-min optimization. Therefore, we must have $\mathbf{f}_1^{\text{new}\dagger} \mathcal{H}_{21} \mathbf{f}_1^{\text{new}} = 0$, which completes the proof of Lemma 1. ■

Now we proceed to prove Theorem 1. Lemma 1 reveals that the updated beamforming vector for user 1 should satisfy $\mathbf{f}_1^{\text{new}\dagger} \mathcal{H}_{21} \mathbf{f}_1^{\text{new}} = 0$. This also implies that it is impossible for the SINR of user 2 to improve over $\gamma^*$ and any interference to user 1 will lead to the result that its achieved SINR becomes smaller than $\gamma^*$. As a by-product, it can be seen that $\mathbf{f}_1^{*\dagger} \mathcal{H}_{21} \mathbf{f}_1^* = 0$.

Based on these observations, if we denote the beamforming update as

$$\Delta \mathbf{f}_1 = \sqrt{p_1^{\text{new}}} \mathbf{f}_1^{\text{new}} - \sqrt{p_1^*} \mathbf{f}_1^*, \tag{36}$$

we must have $\mathbf{h}_{21}^\dagger \Delta \mathbf{f}_1 = 0$. Further, it was shown in [11] that the Pareto optimal beamforming vector for any user always can be expressed as a linear combination of the channels between its serving BS and all the users. Following a similar procedure, it can be proved that the max-min optimal beamforming vector $\mathbf{f}_k^*$ has the same property. That is, they both can be formulated as

$$\mathbf{f}_1 = \zeta_{21} \mathbf{h}_{21} + \zeta_{11} \mathbf{h}_{11} = \tilde{\zeta}_{21} \mathbf{h}_{21} + \zeta_{11} \mathbf{h}_1^{\text{ZF}}, \tag{37}$$

where $\tilde{\zeta}_{21} = \zeta_{21} + \zeta_{11} \frac{\mathbf{h}_{21}^\dagger \mathbf{h}_{11}}{\|\mathbf{h}_{21}\|}$. It follows from $\mathbf{f}_1^\dagger \mathcal{H}_{21} \mathbf{f}_1 = 0$ that $\tilde{\zeta}_{21} = 0$. This means that the beamforming update should be restricted in the direction of $\mathbf{h}_1^{\text{ZF}}$. Thus, the updated unit-norm beamforming vector $\mathbf{f}_1^{\text{new}}$ and the updated transmit power $p_1^{\text{new}}$ can be written as

$$\sqrt{p_1^{\text{new}}} \mathbf{f}_1^{\text{new}} = \sqrt{p_1^*} \mathbf{f}_1^* + b_1 \mathbf{h}_1^{\text{ZF}}, \tag{38}$$

where $b_1$ is a complex scalar. For the maximum Pareto improvement, the optimal power allocation $p_1^{\text{new}}$ and the scaling factor $b_1$ can be determined by solving

$$\max_{\substack{b_1 \in \mathbb{C} \\ 0 < p_1^{\text{new}} \leq P_{\text{BS}}}} \left| \sqrt{p_1^{\text{new}}} \mathbf{h}_{11}^\dagger \mathbf{f}_1^{\text{new}} \right|^2 \quad \text{s.t.} \quad \|\mathbf{f}_1^{\text{new}}\| = 1. \tag{39}$$

Based on the triangle inequality, it can be readily proved that the proposed beamforming update (7) and a full power transmission achieve the maximum rate improvement over the balanced solution $(r^*, r^*)$, i.e., reach the Pareto optimality with max-min fairness. This concludes the proof.

## REFERENCES

[1] P. Marsch and G. Fettweis, "On multicell cooperative transmission in backhaul-constrained cellular systems," *Ann. Telecommun.*, vol. 63, pp. 253–269, 2008.

[2] A. Papadogiannis, D. Gesbert, and E. Hardouin, "A dynamic clustering approach in wireless networks with multi-cell cooperative processing," in Proc. *IEEE Int. Conf. Commun.*, pp. 4033–4037, 2008.

[3] S. Kaviani and W. A. Krzymien, "multicell scheduling in network MIMO," in Proc. *IEEE Int. Global Commun. Conf.*, 2010.

[4] C. T. K. Ng and H. Huang, "Linear precoding in cooperative MIMO cellular networks with limited coordination clusters," *IEEE J. Select. Areas Commun.*, vol. 28, no. 9, pp. 1446–1454, Dec. 2009.

[5] 3GPP R1-090140, "Clustering for CoMP transmission," Nortel.

[6] J. D. Herdtner and E. K. P. Chong, "Analysis of a class of distributed asynchronous power control algorithms for cellular wireless systems," *IEEE J. Select. Areas Commun.*, vol. 18, no.3, pp. 436–446, Mar. 2000.

[7] R. Chen, J. G. Andrews, R. W. Heath Jr., and A. Ghosh, "Uplink power control in multi-cell spatial multiplexing wireless systems," *IEEE Trans. Wireless Commun.*, vol. 6. no. 7, pp. 2700–2711, Jul. 2007.

[8] S. Jing, D. Tse, J. Soriaga, J. Hou, J. Smee, and R. Padovani, "Multicell downlink capacity with coordinated processing," *EURASIP J. Wireless Commun. Net.*, ID. 586878, 2008.

[9] W. Choi and J. G. Andrews, "The capacity gain from intercell scheduling in multi-antenna systems," *IEEE Trans. Wireless Commun.*, vol. 7, no. 2, pp. 714–725, Feb. 2008.

[10] E. Larsson and E. Jorswieck, "Competition versus cooperation on the MISO interference channel," *IEEE J. Select. Areas Commun.*, vol. 26, pp. 1059–1069, 2008.

[11] E. Jorswieck, E. Larsson, and D. Danev, "Complete characterization of the Pareto boundary for the MISO interference channel," *IEEE Trans. Sig. Proc.*, vol. 56, pp. 5292–5296, 2008.

[12] O. Simeone, O. Somekh, H. V. Poor, and S. Shamai (Shitz), "Downlink multicell processing with limited backhaul capacity," *EURASIP J. Advances Sig. Proc. Special Issue on Multiuser MIMO Transmission with Limited Feedback, Cooperation, and Coordination*, Article ID: 840814, 10 pages, 2009.

[13] A. Papadogiannis, E. Hardouin, and D. Gesbert, "Decentralising multicell cooperative processing on the downlink: A novel robust framework," *EURASIP J. Wireless Commun. Net. Special Issue on Broadband Wireless Access*, Article ID: 890685, 10 pages, 2009.

[14] B. L. Ng, J. Evans, S. Hanly, and D. Aktas, "Distributed downlink beamforming with cooperative base stations," *IEEE Trans. Info. Theory*, vol. 54, pp. 5491–5499, 2008.

[15] E. Björnson, R. Zakhour, D. Gesbert, and B. Ottersten, "Cooperative multicell precoding: Rate region characterization and distributed strategies with instantaneous and statistical CSI," *IEEE Trans. Sig. Proc.*, vol. 58, no. 8, pp. 4298–4310, Aug. 2010.

[16] H. Dahrouj and W. Yu, "Coordinated beamforming for the multi-cell multi-antenna wireless system," in Proc. *Conf. Info. Sciences and Sys.*, 19-21 Mar. 2008, Princeton, NJ, USA.

[17] Y. Huang, G. Zheng, M. Bengtsson, K. K. Wong, L. Yang, and B. Ottersten, "Distributed multicell beamforming design with limited intercell coordination," *IEEE Trans. Sig. Proc.*, vol. 59, no. 2, pp. 728–738, Feb. 2011.

[18] B. O. Lee, H. W. Je, I. Sohn, O.-S. Shin, and K. B. Lee, "Interference-aware decentralized precoding for multicell MIMO TDD systems," in Proc. *IEEE Global Commun. Conf.*, pp. 1–5, 30 Nov.-4 Dec. 2008, New Orleans, LA, USA.

[19] N. Hassanpour, J. Smee, J. Hou, and J. Soriaga, "Distributed beamforming based on signal-to-caused-interference ratio," in Proc. *IEEE Int. Sym. Spread Spectrum Tech. and App.*, pp. 405–410, 25-28 Aug. 2008, Bologna, Italy.

[20] M. Bengtsson and B. Ottersten, "Optimal and suboptimal transmit beamforming," in *Handbook of Antennas in Wireless Commun.*, L. C. Godara, Ed., CRC Press, Boca Raton, USA, Aug. 2001.

[21] R1-111282, "Performance evaluation of CoMP JT for Scenario 2," *3GPP TSG RAN WG1 Meeting #65*, Samsung, May 2011.

[22] R1-111290, "CoMP Phase 1 Evaluation Results," *3GPP TSG RAN WG1 Meeting #65*, ZTE, May 2011.

[23] R1-111277, "CoMP JT evaluation for Phase I Homogenous Deployment," *3GPP TSG RAN WG1 Meeting #65*, Texas Instruments, May 2011.

[24] Z. Han, Z. Ji, and K. J. R. Liu, "Fair multiuser channel allocation for OFDMA networks using Nash bargaining solutions and coalitions," *IEEE Trans. Commun.*, vol. 53, no. 8, pp. 1366–1376, Aug. 2005.

[25] J. E. Suris, L. A. DaSilva, Z. Han, A. B. MacKenzie, and R. S. Komali, "Asymptotic optimality for distributed spectrum sharing using bargaining solutions," *IEEE Trans. Wireless Comm.*, vol. 8, no. 10, pp. 5225–5237, Oct. 2009.

[26] A. Wiesel, Y. C. Eldar, and S. Shamai, "Linear precoding via conic optimization for fixed MIMO receivers," *IEEE Trans. Sig. Proc.*, vol. 54, no. 1, pp. 161–176, Jan. 2006.

[27] M. Grant and S. Boyd, "CVX: Matlab software for disciplined convex programming," Available [Online] http://stanford.edu/∼boyd/cvx, Jun. 2009.

[28] W. Yu and T. Lan, "Transmitter optimization for the multi-antenna downlink with per-antenna power constraints," *IEEE Trans. Sig. Proc.*, vol. 55, no. 6, pp. 2646–2660, Jun. 2007.

[29] M. Schubert and H. Boche, "Solution of the multiuser downlink beamforming problem with individual SINR constraints," *IEEE Trans. Veh. Tech.*, vol. 53, no. 1, pp. 18–28, Jan. 2004.

[30] R. Horn and C. Johnson, *Matrix analysis*. Cambridge University Press, 1985.

[31] R. D. Yates, "A framework for uplink power control in cellular radio systems," *IEEE J. Select. Areas Commun.*, vol. 13, no. 7, pp.1341–1347, Sep. 1995.

[32] M. B. Shenouda and T. N. Davidson, "On the design of linear transceivers for multiuser systems with channel uncertainty," *IEEE J. Select. Areas Commun.*, vol. 26, no. 6, pp. 1015–1024, Aug. 2008.

[33] X. Zhang, D. P. Palomar, and B. Ottersten, "Statistically robust design of linear MIMO transceivers," *IEEE Trans. Sig. Proc.*, vol. 56, no. 8, pp. 3678–3689, Aug. 2008.

[34] T. E. Bogale, L. Vandendorpe, and B. K. Chalise, "Robust transceiver optimization for downlink coordinated base station systems: distributed algorithm," *IEEE Trans. Sig. Proc.*, vol. 60, no. 1, pp. 337–350, Jan. 2012.

[35] A. Tajer, N. Prasad, and X. Wang, "Robust linear precoder design for multi-cell downlink transmission," *IEEE Trans. Sig. Proc.*, vol. 59, no. 1, pp. 235–251, Jan. 2011.

[36] N. Vucic and H. Boche, "Robust QoS-constrained optimization of downlink multiuser MISO systems," *IEEE Trans. Sig. Proc.*, vol. 57, no. 2, pp. 714–725, Feb. 2009.

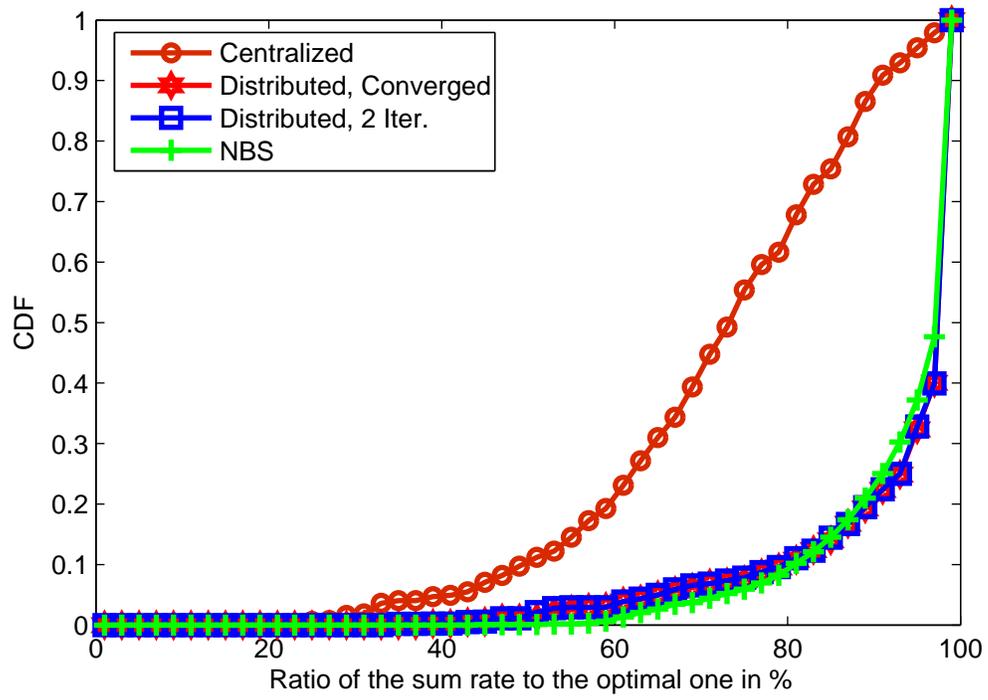

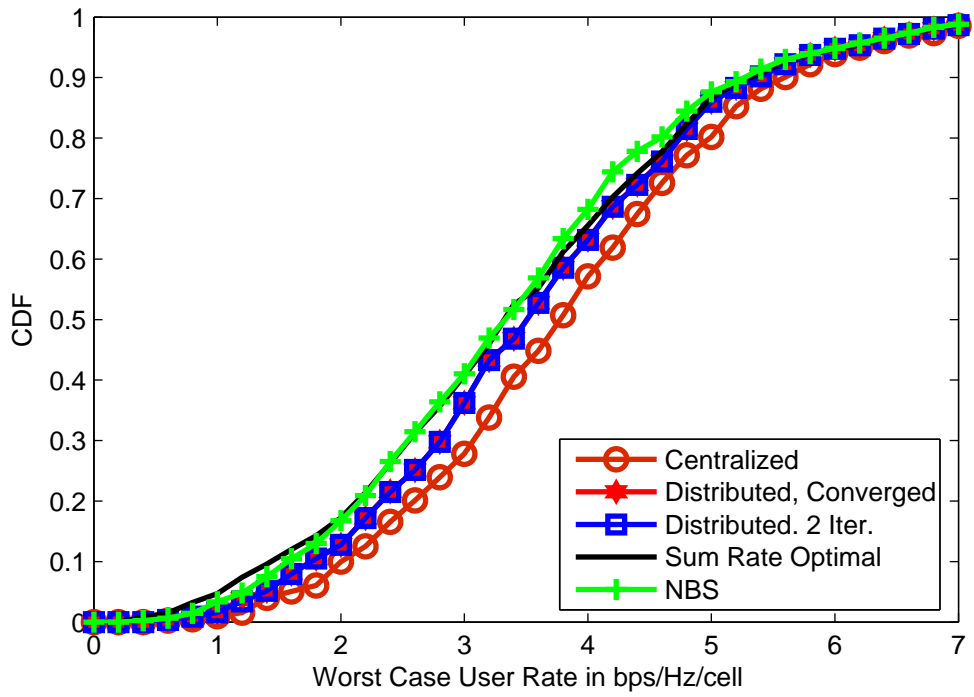